\documentclass[twocolumn,showkeys,aps,prb,showpacs]{revtex4-1}
\UseRawInputEncoding
\usepackage{graphicx}
\usepackage[CJKbookmarks,dvipdfm,colorlinks,linkcolor=red,citecolor=blue]{hyperref}

\begin{document}

\title{Piezoelectric ferromagnetism in Janus monolayer YBrI: a first-principle prediction}

\author{San-Dong Guo$^{1}$, Meng-Xia Wang$^{1}$, Yu-Ling Tao$^{1}$  and  Bang-Gui Liu$^{2,3}$}
\affiliation{$^1$School of Electronic Engineering, Xi'an University of Posts and Telecommunications, Xi'an 710121, China}
\affiliation{$^2$ Beijing National Laboratory for Condensed Matter Physics, Institute of Physics, Chinese Academy of Sciences, Beijing 100190, People's Republic of China}
\affiliation{$^3$School of Physical Sciences, University of Chinese Academy of Sciences, Beijing 100190, People's Republic of China}
\begin{abstract}
Coexistence of intrinsic ferromagnetism and piezoelectricity, namely piezoelectric ferromagnetism (PFM), is crucial to advance multifunctional spintronic technologies.
In this work,  we demonstrate that Janus monolayer YBrI is a PFM, which is dynamically, mechanically  and thermally stable. Electronic correlation effects on physical properties of YBrI are investigated by using  generalized gradient approximation plus $U$ (GGA+$U$) approach.
For out-of-plane magnetic anisotropy, YBrI is a ferrovalley (FV) material, and the valley splitting is larger than 82 meV in considered $U$ range.
 The anomalous valley Hall effect (AVHE) can be achieved under an in-plane electric field. However, for in-plane  magnetic anisotropy, YBrI is a common ferromagnetic (FM) semiconductor.  When considering intrinsic magnetic anisotropy, the  easy axis of YBrI is always in-plane with magnetic anisotropy energy (MAE) from 0.309 meV to 0.237 meV ($U$=0.0 eV to 3.0 eV). However, the
magnetization can be adjusted from the in-plane to off-plane
direction by external magnetic field, and then  lead to the occurrence of valley
polarization.  Moreover, missing centrosymmetry along with mirror symmetry breaking results
in both in-plane and out-of-plane piezoelectricity in YBrI monolayer. At a typical $U$=2.0 eV, the $d_{11}$ is predicted to be -5.61 pm/V, which  is higher than or  compared with ones of  other two-dimensional (2D) known materials. The electronic and piezoelectric properties of YBrI  can be effectively tuned by applying a biaxial strain. For example, tensile strain can enhance valley splitting and $d_{11}$ (absolute value). The predicted  Curie temperature of YBrI is higher than those of  experimentally synthesized 2D ferromagnetic
materials $\mathrm{CrI_3}$  and $\mathrm{Cr_2Ge_2Te_6}$.
Our findings of these distinctive properties  could pave the way to design  multifunctional
 spintronic  devices, and bring  forward a new perspective for constructing 2D materials.

\end{abstract}
\keywords{Ferromagnetism, Piezoelectricity, 2D materials~~~~~~~~~~~~~~~~~~~~~~~Email:sandongyuwang@163.com}

\maketitle

\section{Introduction}
2D materials have attracted much attention due to their unique properties and
potential applications in future nanodevices\cite{f1}.
The reduction in dimensionality of 2D
materials results in that  inversion symmetry
is often eliminated, which
allows these materials to become piezoelectric. The $\mathrm{MoS_2}$ monolayer is a classical 2D piezoelectric material, which  is firstly predicted theoretically\cite{f2}, and then is confirmed experimentally\cite{f3}.
Magnetism in 2D systems is one of the most fascinating
properties of materials, and long-range magnetic order should be prohibited according to the
Mermin-Wagner theorem\cite{f3-1}. However, the intrinsic long-range FM order in 2D  semiconductors such as $\mathrm{Cr_2Ge_2Te_6}$ and
$\mathrm{CrI_3}$\cite{f3-2,f3-3} has been achieved experimentally, which is due to  spin-orbital coupling (SOC)-induced magnetic anisotropy.
Valley is emerging as a new degree of freedom, which  can
be used to manipulate  information, namely  valleytronics\cite{f4,f5}.
 To achieve valley application, the valley polarization should be induced.
The  2D magnetic material with special crystal structure  provides an opportunity to
realize spontaneous valley polarization, which is
known as FV  material\cite{f6}.
\begin{figure*}
  \includegraphics[width=12cm]{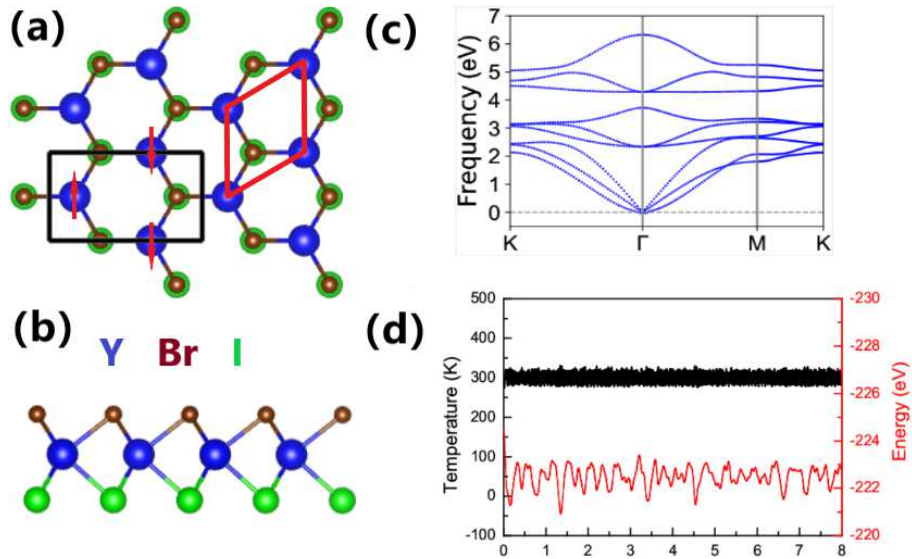}
  \caption{(Color online)For YBrI monolayer,  (a): top view and (b): side view of  crystal structure. The primitive (rectangle supercell) cell is
   marked by red (black) lines, and the AFM configuration is shown with arrows in (a). (c):the phonon dispersion curves. (d):the temperature and total energy fluctuations as a function of simulation time at 300 K.}\label{t0}
\end{figure*}

\begin{figure}
   \includegraphics[width=8cm]{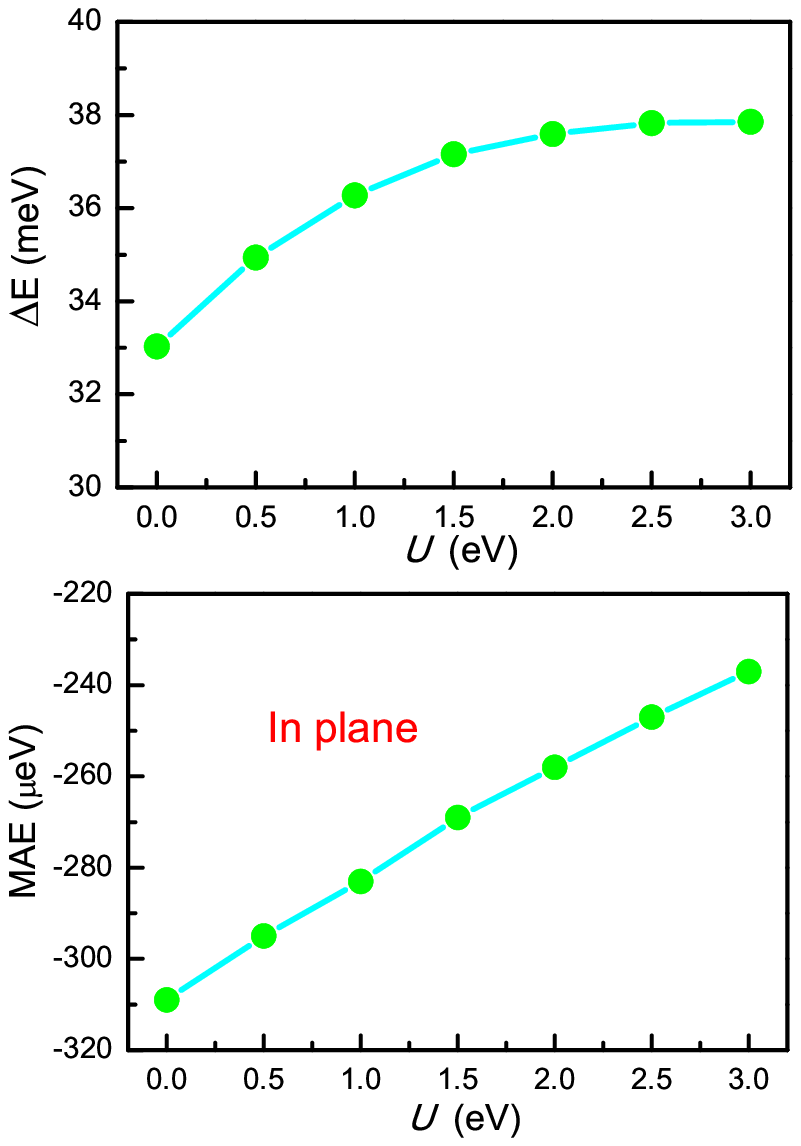}
  \caption{(Color online)For monolayer YBrI, the energy differences  between  AFM and FM ordering and MAE  as a function of $U$.}\label{u-em}
\end{figure}
\begin{figure*}
   \includegraphics[width=16cm]{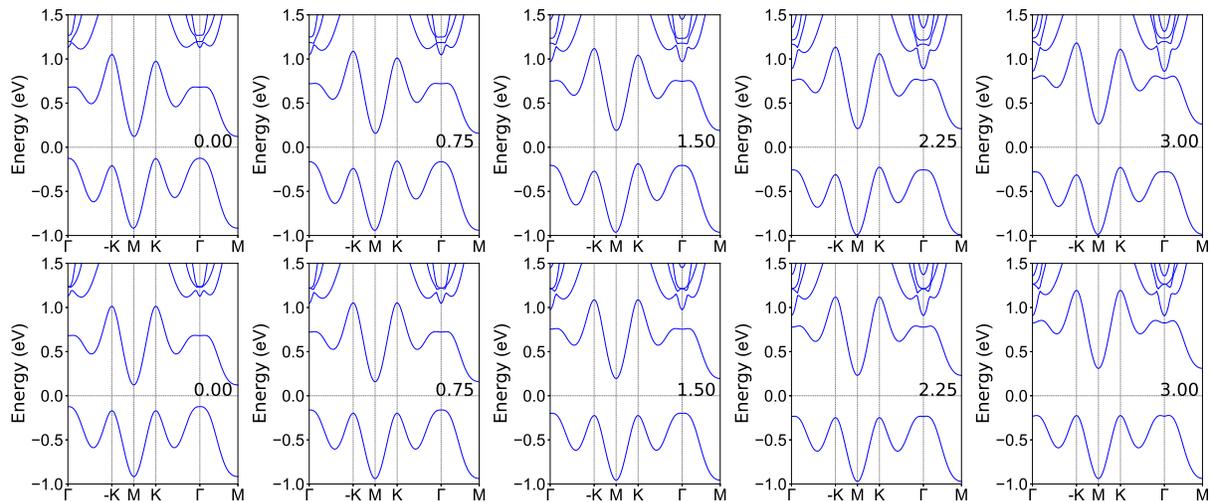}
  \caption{(Color online) The energy band structures of YBrI monolayer with out-of-plane (Top) or in-plane (Bottom)  magnetic anisotropy  by using GGA+SOC at some representative $U$ values. }\label{u-band}
\end{figure*}
\begin{figure}
   \includegraphics[width=8cm]{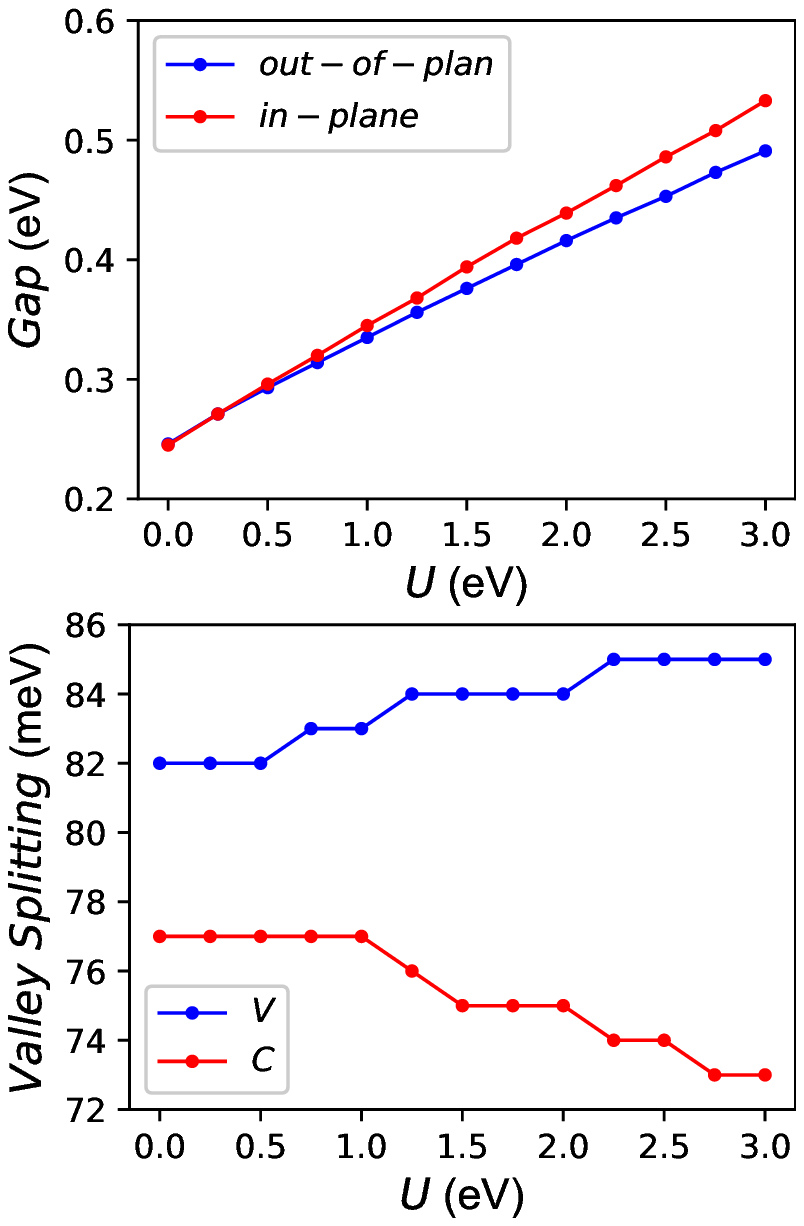}
  \caption{(Color online)For  monolayer YBrI, the gap (including out-of-plane and in-plane  magnetic anisotropy) and valley splitting (including  bottom conduction ($C$) and top valence ($V$) bands)  as a function of $U$.}\label{u-gap}
\end{figure}

Compared with 2D materials with individual  piezoelectric or magnetic or valley  property,  2D multifunctional materials may give rise to
unprecedented opportunities for intriguing physics, and produce  novel device applications\cite{f7}.
Coexistence of intrinsic ferromagnetism and piezoelectricity has been reported in some 2D systems\cite{q15-0,q15,q15-1,q15-2}.
The strong out-of-plane piezoelectricity is predicted in PFM  $\mathrm{CrBr_{1.5}I_{1.5}}$ monolayer, and the $d_{31}$ is larger than 1.0 pm/V\cite{q15-1}.
The FV materials possess magnetism, and lack spatial inversion-symmetry\cite{f6}, which make these materials to be PFMs.  To realize AVHE, a possible way  is proposed  in FV monolayer $\mathrm{GdCl_2}$ by piezoelectric effect, not an  external
electric field, namely piezoelectric anomalous valley Hall effect (PAVHE)\cite{gsd1}.
The piezoelectricity and FV properties in Janus monolayer VClBr are investigated, and  a giant in-plane piezoelectricity is predicted\cite{f8}.

Recently,   $\mathrm{YX_2}$(X=I, Br and Cl) monolayers are predicted to be  dynamically
and thermally stable 2D FM semiconductors, and the large spontaneous valley polarizations  are also
reported\cite{f9}. However,  magnetic anisotropy is not considered, and an out-of-plane magnetic anisotropy  is very key to produce spontaneous valley polarization\cite{f10,f11}.  For monolayer FeClF,  increasing electron correlation effects can induce coexistence of FV and quantum anomalous Hall (QAH) state with fixed out-of-plane magnetic anisotropy, but the spontaneous valley polarization will disappear with varied correlation strength for in-plane case\cite{f11}.

In this work, taking Janus monolayer YBrI as a example as derivative of  $\mathrm{YX_2}$(X=I, Br and Cl), we investigate its magnetic, electronic and piezoelectric properties. Our results show that monolayer YBrI is  dynamically, mechanically  and thermally stable, and is intrinsically a common FM semiconductor due to inherent in-plane magnetic anisotropy.  To achieve FV properties, an  external magnetic field is needed to adjust magnetization form in-plane  to off-plane
direction. With breaking mirror or inversion
symmetry along out-of-plane orientation,  both in-plane and out-of-plane piezoelectricity can be induced in YBrI monolayer with a imposed uniaxial in-plane strain.
 At a typical $U$=2.0 eV, the predicted $d_{11}$ (-5.61 pm/V)   is higher than or  compared with ones of  other 2D known materials\cite{f2,f12}.  Calculated results show that the strain can effectively tune electronic and piezoelectric properties of YBrI.  The predicted  Curie temperature of YBrI is higher than those of  experimentally synthesized ferromagnets $\mathrm{CrI_3}$  and $\mathrm{Cr_2Ge_2Te_6}$\cite{f3-2,f3-3}. Our works imply that the Janus monolayer YBrI is a potential multifunctional material with possible applications in spintronics and piezoelectronics.

\begin{figure*}
  \includegraphics[width=16cm]{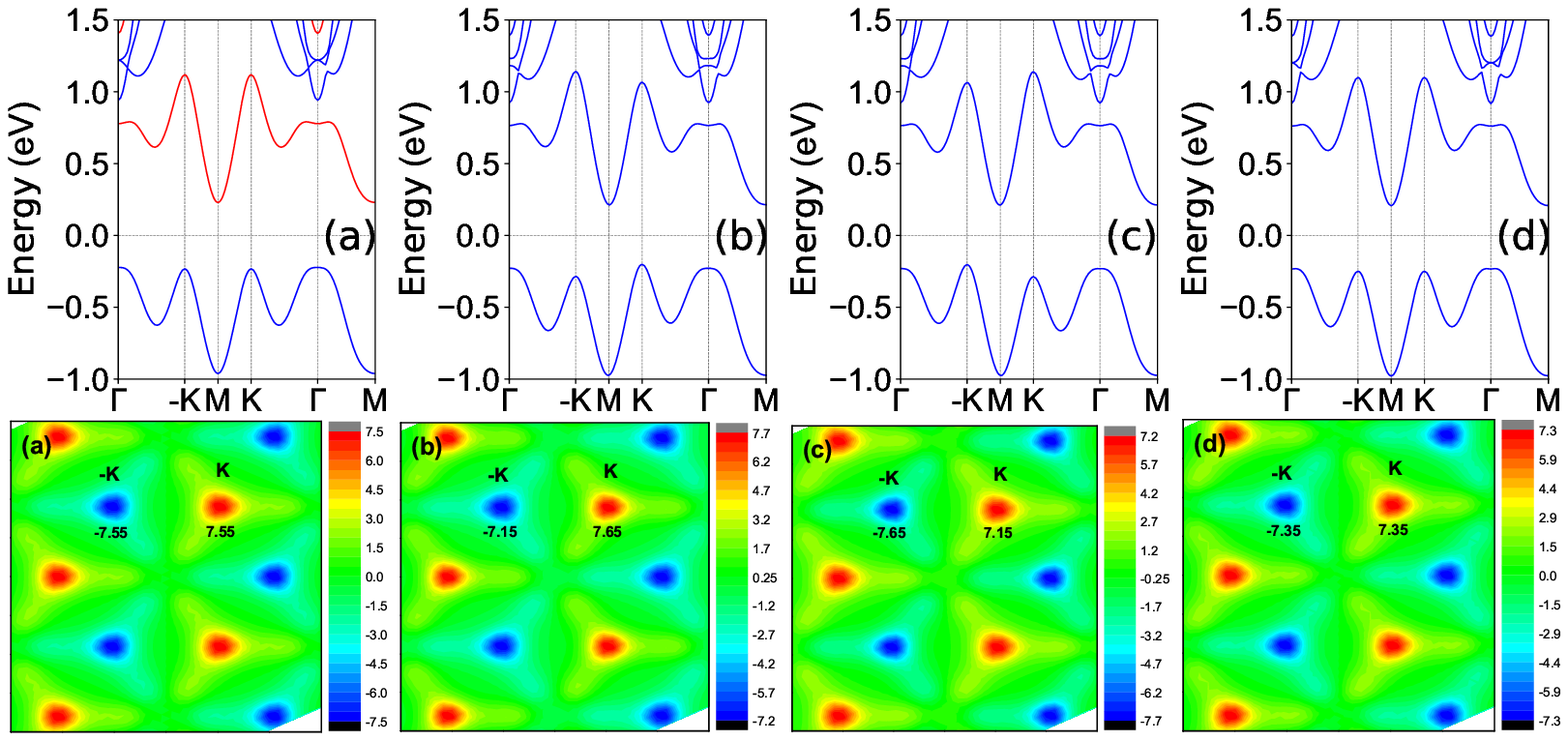}
\caption{(Color online)The band structures (Top) and  Berry curvature distributions (Bottom) of  monolayer  YBrI (a) without SOC; (b), (c) and (d) with SOC for magnetic moment of Y along the positive $z$, negative $z$, and positive $x$ direction, respectively.  In (a), the blue (red) lines represent the band structure in the spin-up (spin-down) direction.}\label{band-z}
\end{figure*}
\begin{figure}
   \includegraphics[width=8cm]{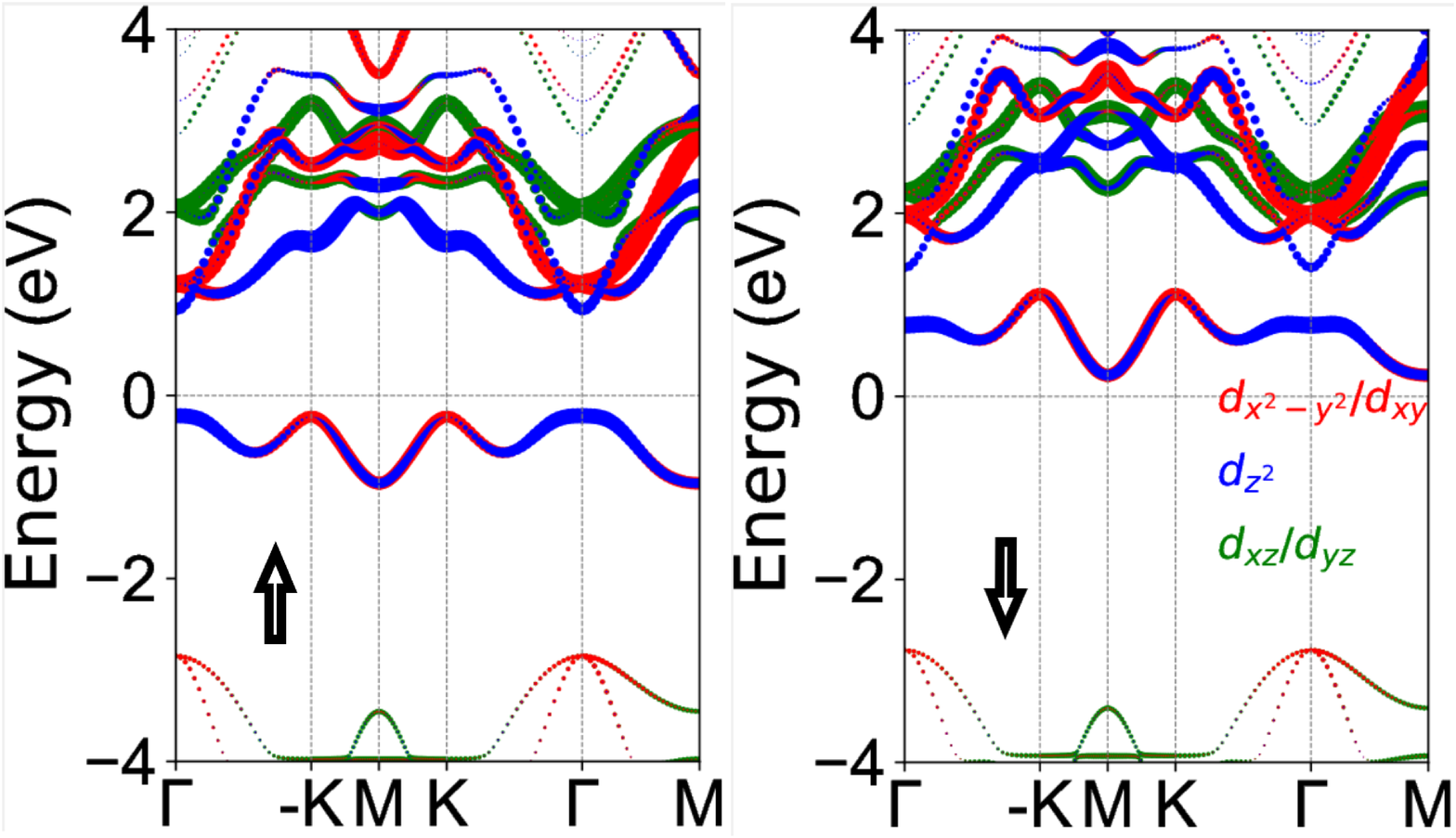}
  \caption{(Color online)For  monolayer YBrI, Y-$d$-orbital characters of energy bands  by using GGA+$U$ ($U$$=$2.0 eV), including spin-up and spin-down directions.}\label{p}
\end{figure}

\section{Computational detail}
Our first-principles calculations are performed by means of
the density functional theory (DFT)\cite{1} with projector-augmented-wave (PAW) potential and
plane-wave basis, as implemented in the VASP code\cite{pv1,pv2,pv3}.
 The  exchange-correlation functional is tackled by GGA of Perdew, Burke and  Ernzerhof (PBE-GGA)\cite{pbe}.
A  kinetic energy cutoff of 500 eV,  a force convergence criterion of less than 0.0001 $\mathrm{eV.{\AA}^{-1}}$, and a total energy  convergence criterion of  $10^{-8}$ eV are adopted to attain reliable results.
The GGA+$U$ scheme by the
rotationally invariant approach proposed by Dudarev et al\cite{u} is adopted to account for the strong correction effect for the localized Y-4$d$ electrons.  A vacuum region of more than 18 $\mathrm{{\AA}}$ along the $z$ direction is used
to eliminate the image interaction between adjacent layers. SOC effect is fully considered in the calculations of both
band structure and MAE.

 The phonon dispersions with FM ground state are calculated
 through the direct supercell method with the 5$\times$5$\times$1 supercell, as implemented in Phonopy code\cite{pv5}.
Ab initio
molecular dynamics (AIMD) simulation over a 4$\times$4$\times$1
supercell is performed  in the canonical ensemble for 8 ps with a
time step of 1.0 fs, where the temperature is controlled at
300 K.
The Curie temperature is
estimated  by Monte Carlo (MC) simulations, as implemented in Mcsolver code\cite{mc}.
 The Berry curvatures of YBrI monolayer
are calculated directly from the calculated
wave functions,  based on Fukui's
method\cite{bm},  as implemented in the VASPBERRY code\cite{bm1,bm2}.
The elastic stiffness tensor  $C_{ij}$  and piezoelectric stress tensor $e_{ij}$ are carried out by using strain-stress relationship and density functional perturbation theory (DFPT) method\cite{pv6}. The 2D  $C^{2D}_{ij}$/$e^{2D}_{ij}$
have been renormalized by   $C^{2D}_{ij}$=$L_z$$C^{3D}_{ij}$/$e^{2D}_{ij}$=$L_z$$e^{3D}_{ij}$, in which  $L_z$ is  the length of unit cell along $z$ direction.
 The Brillouin zone (BZ) integration is sampled by using a
21$\times$21$\times$1 Monkhorst-Pack grids for self-consistent calculations   and  $C_{ij}$, and  12$\times$21$\times$1 k-points for  FM/ntiferromagnetic (AFM) energy and  $e_{ij}$.

\section{Structure and stability}
In analogy to  monolayer MoSSe\cite{e1,e2}, YBrI consists of Br-Y-I sandwich layer. Each Y
atom is surrounded by three Br and three I atoms, which  forms   a triangular prism.  The  top and side views of crystal structure of YBrI are shown in \autoref{t0}, which can be constructed  by replacing one of two  Br/I   layers with I/Br atoms in $\mathrm{YBr_2}$/$\mathrm{YI_2}$.
 As a result, the inequivalent bond lengths between Y-Br and Y-I reduce the symmetry from  $P\bar{6}m2$ to $P3m1$ and
break the mirror symmetry, compared with monolayer $\mathrm{YBr_2}$/$\mathrm{YI_2}$.
For YBrI monolayer, the broken  mirror symmetry  leads to  both in-plane and vertical piezoelectric polarizations, and only in-plane piezoelectric polarization exists in $\mathrm{YBr_2}$/$\mathrm{YI_2}$.
It has been reported that the electronic
properties of monolayer  FeClF, which has the same crystal structure with YBrI,  are sensitive to electronic correlation effects\cite{f11}. To study the electronic correlation effect
of YBrI, the lattice constants $a$ is optimized at different $U$ (0-3 eV), as shown in FIG.1 of electronic supplementary information (ESI). It is found that the $a$ almost linearly increases with increasing $U$, and the change only is 0.054 $\mathrm{{\AA}}$.  The energy difference of  AFM (see \autoref{t0}) and FM configurations  as a function of $U$ is calculated to
determine the ground state of YBrI, as is plotted in \autoref{u-em}. In considered $U$ range, YBrI is always a FM ground state, and increasing $U$ can strengthen FM interaction.

To evaluate the stability of YBrI,
the phonon dispersion, molecular dynamics and elastic
constant calculations are carried out by using GGA method.
The phonon spectrum is calculated to confirm the dynamic
stability of YBrI (see \autoref{t0}). Due to containing  three atoms in primitive cell,  the phonon
dispersion possesses nine branches, including three acoustic and six
optical modes.  No imaginary frequencies can be observed in YBrI monolayer,  indicating its dynamic stability.
To corroborate the thermal stability, the evolution of temperature and total
energy vs time are calculated using AIMD at room temperature, which is plotted in \autoref{t0}.
During the
simulation period, with increasing
time, the frameworks of YBrI are well preserved with   little fluctuations of temperature and total
energy,   which confirms its thermal stability. Due to $P3m1$ space group,  using Voigt notation,
the elastic tensor of YBrI  can be reduced into:
\begin{equation}\label{pe1-4}
   C=\left(
    \begin{array}{ccc}
      C_{11} & C_{12} & 0 \\
     C_{12} & C_{11} &0 \\
      0 & 0 & (C_{11}-C_{12})/2 \\
    \end{array}
  \right)
\end{equation}
The independent $C_{11}$ and $C_{12}$  of YBrI are 41.30 $\mathrm{Nm^{-1}}$ and 12.23 $\mathrm{Nm^{-1}}$, which  fulfill these conditions of Born  criteria of mechanical stability\cite{ela}: $C_{11}$$>$0 and $C_{11}-C_{12}$$>$0,   confirming its  mechanical stability.

\begin{figure}
   \includegraphics[width=6cm]{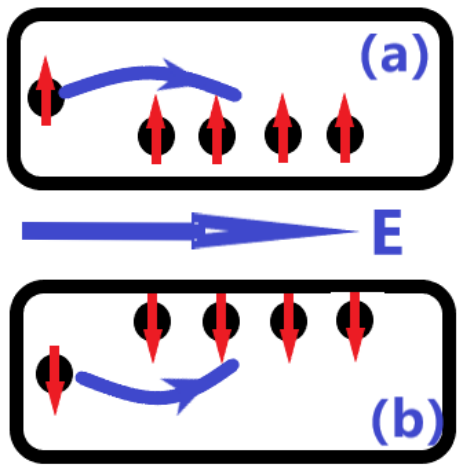}
  \caption{(Color online) Schematic
AVHE for the hole-doped YBrI monolayer at the K (a) and -K (b) valleys under an in-plane electric field $E$. The  arrows
stand for the spin-up or spin-down states.}\label{av}
\end{figure}

\section{correlation effects}
Electronic correlation and magnetic anisotropy have important effects on electronic states of  some 2D materials\cite{f10,f11}.
For FeClF monolayer,  increasing electron correlation effects with fixed out-of-plane magnetic anisotropy can induce the FV to half-valley-metal (HVM) to QAH to HVM to FV transitions\cite{f11}.
However, for in-plane magnetic anisotropy, these  novel electronic states  will disappear.
The YBrI has the same crystal structure with FeClF. So, it is very necessary for electronic state calculations of YBrI to consider electronic correlation and magnetic anisotropy.

Firstly, the out-of-plane magnetic anisotropy is assumed with varied $U$, which  maintains the horizontal mirror symmetry, and breaks all
possible vertical mirrors.  The evolutions of electronic band
structures as a function of $U$ are studied by GGA+SOC, and  the energy band structures at  representative $U$ values are plotted in \autoref{u-band}.
It is clearly seen that they all are indirect gap semiconductors in considered $U$ range, and a spontaneous
valley splitting can be  observed. These mean that YBrI is a FV material. The gap and  valley splitting defined as the energy difference between -K and K valleys (absolute value) as a function of $U$ are plotted in \autoref{u-gap}. For FeClF monolayer, the gap closes,
reopens, and then closes\cite{f11}. However, the gap  increases monotonically with increasing $U$ for YBrI monolayer, which means that no  topological phase transition is produced. For top valence band,  the general tendency of   valley splitting (absolute value) increases with increasing $U$, while the reduced tendency is observed for that of bottom conduction band. It should point out that only valley polarization of top valence band can  be applied in practice, and the  valley splitting (absolute value) is larger than 82 meV in considered $U$ range.  This value is larger than those of
many predicted FV materials such as  $\mathrm{VSe_2}$,  $\mathrm{LaBr_2}$ and $\mathrm{VSi_2N_4}$(33 meV-78.2 meV)\cite{f6,q13,q13-1,q14}.

With fixed in-plane  magnetic anisotropy,   the energy band structures at  representative $U$ values are also shown in \autoref{u-band}, and the gaps vs $U$ are  plotted \autoref{u-gap}. It is clearly seen that YBrI is still an indirect gap semiconductor in considered $U$ range, and the gap increases monotonically, when $U$ increases. However, the -K and
K valleys  keep degenerate, which means that no spontaneous
valley splitting can be observed. Similar results can be found in some 2D materials\cite{f10,f11}.
For example, for FeClF\cite{f11}, varied $U$ produces  no special QAH states and prominent valley polarization, which is very different from those with fixed out-of-plane magnetic anisotropy.

In view of the above discussion, it is very important to confirm intrinsic  magnetic anisotropy of monolayer YBrI, which can be described by MAE.  The MAE can be calculated  by the energy
difference of the magnetization orientation along the (100)
and (001) cases.  The positive values mean out-of-plane magnetic anisotropy, while negative values indicate in-plane one.
The MAE as a function of $U$ is plotted in \autoref{u-em}. It is clearly seen that the MAE of YBrI is always negative, indicating its in-plane magnetic anisotropy. The MAE changes from -0.309 meV to -0.237 meV, and  increasing $U$ can reduce MAE. So, YBrI intrinsically is not a FV material.
However, the FV states can be achieved  by external magnetic field, which can change in-plane magnetic anisotropy to out-of-plane one.

\begin{figure}
   \includegraphics[width=8cm]{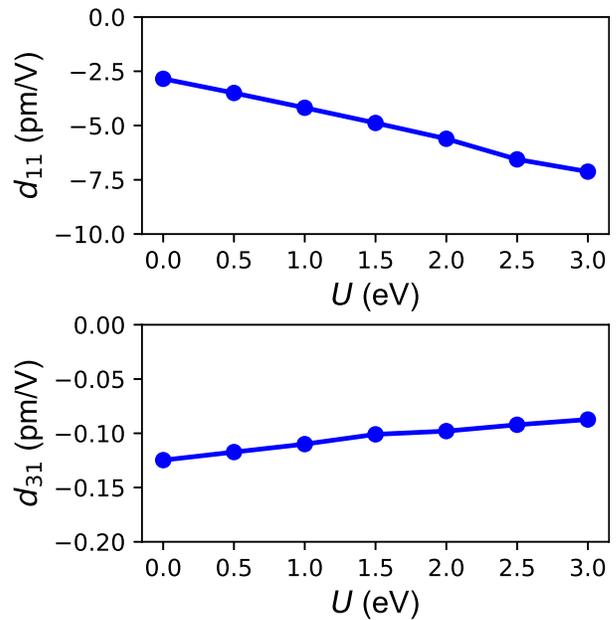}
  \caption{(Color online)For  monolayer YBrI, the piezoelectric strain coefficient $d_{11}$ and $d_{31}$  as a function of $U$.}\label{u-d11}
\end{figure}

Finally, taking a typical $U$=2.0 eV, the electronic states of YBrI are detailedly investigated.  The spin-polarized band structures of monolayer  YBrI by using both GGA and GGA+SOC are plotted in \autoref{band-z}.
According to \autoref{band-z} (a),  a distinct
spin splitting can be observed due to the exchange
interaction, and  YBrI is  a direct band
gap semiconductor (gap of 0.463 eV) with valence band maximum (VBM)/conduction band bottom (CBM)  provided by the spin-up/spin-down.
For both conduction and valence bands, the energies of valleys of -K and K are degenerate. Based on  projected band structure in \autoref{p},
it is clearly seen that the only one electron
 occupies  top $d_{z^2}$-dominated valence band in spin-up direction,  while the other
bottom $d_{z^2}$-dominated conduction band in spin-dn direction is empty.
 This special electron configuration will lead to that
the magnetic moment of each Y atom is expected to be
1 $\mu_B$. \autoref{band-z} (b) shows that the  valley polarization can be induced by SOC.
The valley splitting of top valence band   is 84  meV, which is between ones of  $\mathrm{YBr_2}$ (58 meV) and $\mathrm{YI_2}$ (109 meV)\cite{f9}. The energy of K valley
is higher than one of -K valley, and  the valley polarization can  be
switched by reversing the magnetization direction (see \autoref{band-z} (c)). \autoref{band-z} (b) and (c) show that YBrI is  a direct band
gap semiconductor with gap value of 0.416 eV. When the  magnetization direction of YBrI  is in-plane along $x$ direction (see \autoref{band-z} (d)), no valley polarization can be observed, and it is still  a direct band
gap semiconductor with gap value of 0.439 eV.

According to \autoref{p}, it is found that $d_{x^2-y^2}$/$d_{xy}$ orbitals  dominate -K and K valleys, which is very key to produce valley polarization.
The valley polarization induced by SOC  is mainly due to  the intra-atomic interaction  $\hat{H}^0_{SOC}$ (the interaction
between the same spin states). For out-of-plane magnetization,  $\hat{H}^0_{SOC}$  can be expressed as\cite{f6,v2,v3}:
\begin{equation}\label{m1}
\hat{H}^0_{SOC}=\alpha \hat{L}_z
\end{equation}
in which  $\hat{L}_z$ is the orbital angular moment along $z$ direction, and $\alpha$ is the coupling strength. The resulting energy  at K
and -K valleys can be expressed as:
\begin{equation}\label{m3}
E^\tau=<\phi^\tau|\hat{H}^0_{SOC}|\phi^\tau>
\end{equation}
where $|\phi^\tau>$ means  the orbital
basis for -K and K valleys, and the subscript  $\tau$ represents  valley index ($\tau=\pm1$).
If the -K and K valleys are
mainly from   $d_{x^2-y^2}$/$d_{xy}$ orbitals, the valley splitting $|\Delta E|$
 at -K and K points  can be written as:
\begin{equation}\label{m4}
|\Delta E|=E^{K}-E^{-K}=4\alpha
\end{equation}
If the -K and K valleys are dominated by $d_{z^2}$ orbitals, the valley splitting $|\Delta E|$
 is expressed as:
\begin{equation}\label{m4}
|\Delta E|=E^{K}-E^{-K}=0
\end{equation}
For $d_{x^2-y^2}$/$d_{xy}$-dominated -K/K valley,
when the magnetization orientation is general case, $\Delta E=4\alpha cos\theta$\cite{v3} ($\theta$=0/90$^{\circ}$ means out-of-plane/in-plane direction.). When the magnetocrystalline direction  is along in-plane one, the valley splitting of YBrI will be zero.

\autoref{band-z} also presents the calculated Berry curvature of
YBrI as a contour map in 2D BZ with and without SOC. For all four situations, the opposite
signs around -K and K valleys can be observed. For these situations without valley polarization, Berry curvatures  show  equal magnitudes at -K and K valleys. However, unequal magnitudes of Berry curvatures at -K and K valleys can be observed for those with valley polarization. When reversing
the magnetization from the $z$ to $-z$ direction, the magnitudes
of Berry curvature at -K and K valleys exchange to each
other but their signs remain unchanged.

Under an in-plane longitudinal electric field $E$, anomalous velocity $\upsilon$  of Bloch electrons  is associated with Berry curvature $\Omega(k)$:$\upsilon\sim E\times\Omega(k)$\cite{q9}.
An appropriate hole doping
can move the Fermi level to fall between the -K and K
valleys.  This leads that only one valley has doped holes. When  an in-plane electric field is applied, the  Berry curvature forces
the hole carriers to accumulate on one side of the sample,  giving rise to an
AVHE in monolayer YBrI.  As illustrated in \autoref{av}, when K valley is doped with holes, the positive Berry curvature drives the spin-up holes  to accumulate on the right side of the sample in the
presence of an in-plane electric field. When the magnetization
is reversed,  -K valley is doped with holes, and then the spin-down holes  move
to the left side of the sample due to negative
Berry curvature of -K valley.

\begin{figure}
   \includegraphics[width=8cm]{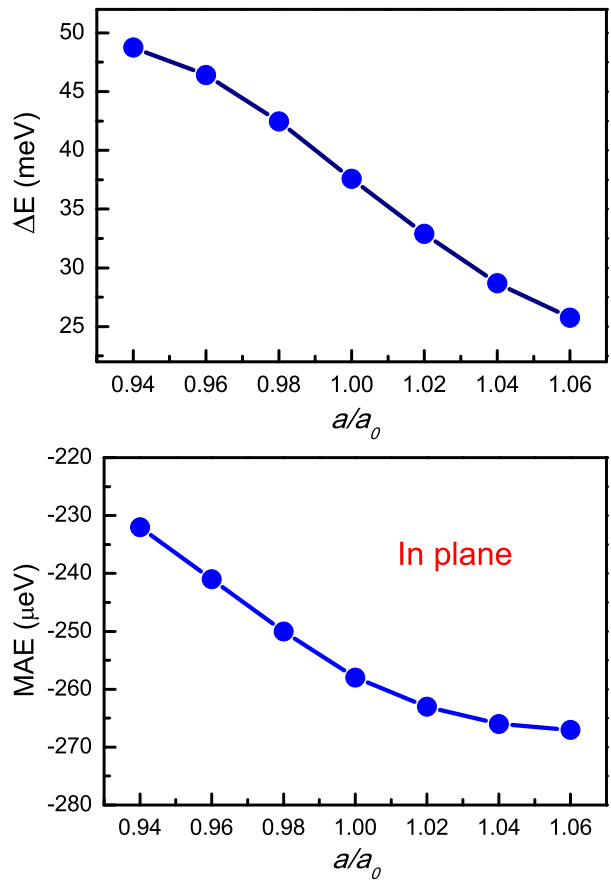}
  \caption{(Color online)For monolayer YBrI, the energy differences  between  AFM and FM ordering and MAE  as a function of $a/a_0$.}\label{s-ae}
\end{figure}

\begin{figure}
   \includegraphics[width=8cm]{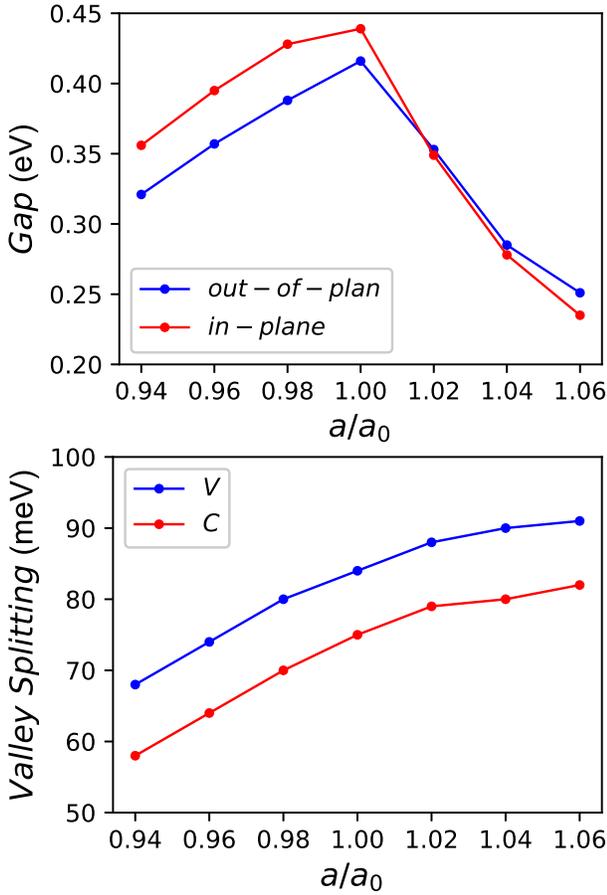}
  \caption{(Color online))For  monolayer YBrI, the gap (including out-of-plane and in-plane  magnetic anisotropy) and valley splitting (including  bottom conduction ($C$) and top valence ($V$) bands)  as a function of $a/a_0$.}\label{s-gap}
\end{figure}

\section{Piezoelectric properties}
The YBrI monolayer lacks  inversion symmetry, and then it is piezoelectric. Moreover, both in-plane and vertical piezoelectric polarizations can be achieved due to special Janus structure, which is the same with Janus  MoSSe monolayer\cite{f18}.
Based on energy band structures above,  Janus YBrI is a FM semiconductor, and   its semiconductor properties is compatible as a practical piezoelectric material  to  prohibit current leakage. These indicate that the piezoelectricity and magnetism  can  coexist in YBrI.

Being analogous to MoSSe monolayer\cite{f18},  by performing symmetry analysis and only considering the in-plane strain and stress, the  piezoelectric stress   and strain tensors of YBrI  with Voigt notation can be reduced into:
 \begin{equation}\label{pe1-1}
 e=\left(
    \begin{array}{ccc}
      e_{11} & -e_{11} & 0 \\
     0 & 0 & -e_{11} \\
      e_{31} & e_{31} & 0 \\
    \end{array}
  \right)
    \end{equation}

  \begin{equation}\label{pe1-2}
  d= \left(
    \begin{array}{ccc}
      d_{11} & -d_{11} & 0 \\
      0 & 0 & -2d_{11} \\
      d_{31} & d_{31} &0 \\
    \end{array}
  \right)
\end{equation}
  When a uniaxial in-plane strain is imposed,   both in-plane and out-of-plane piezoelectric polarization  can be induced ($e_{11}$/$d_{11}$$\neq$0 and $e_{31}$/$d_{31}$$\neq$0). However, with   an applied biaxial in-plane strain,  the
in-plane piezoelectric response will disappear($e_{11}$/$d_{11}$=0), but the out-of-plane piezoelectric polarization still maintains ($e_{31}$/$d_{31}$$\neq$0).
Here, the two independent $d_{11}$ and $d_{31}$ can be derived by $e_{ik}=d_{ij}C_{jk}$:
\begin{equation}\label{pe2}
    d_{11}=\frac{e_{11}}{C_{11}-C_{12}}~~~and~~~d_{31}=\frac{e_{31}}{C_{11}+C_{12}}
\end{equation}

\begin{figure}
   \includegraphics[width=8cm]{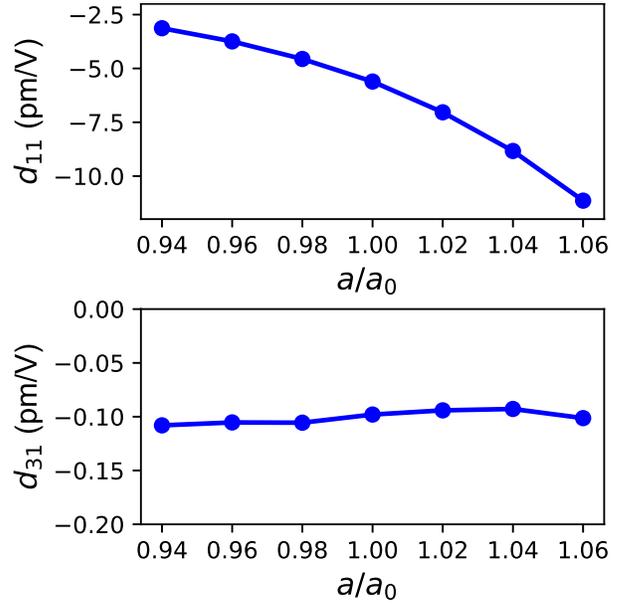}
  \caption{(Color online)For  monolayer YBrI, the piezoelectric strain coefficient $d_{11}$ and $d_{31}$  as a function of $a/a_0$.}\label{s-d11}
\end{figure}

We use orthorhombic supercell (see  \autoref{t0}) as the
computational unit cell  to calculate the  $e_{11}$/$e_{31}$ of YBrI.
The elastic constants  ($C_{11}$, $C_{12}$,  $C_{11}$-$C_{12}$ and $C_{11}$+$C_{12}$) and  piezoelectric  stress  coefficients  ($e_{11}$ and $e_{31}$) along  the ionic  and electronic contributions  as a function of $U$ are shown in FIG.2 and FIG.3 of ESI.
The $d_{11}$/$d_{31}$ vs $U$ are plotted in \autoref{u-d11}. It is found that $C_{11}$, $C_{12}$,  $C_{11}$-$C_{12}$,  $C_{11}$+$C_{12}$, $e_{31}$ and $d_{31}$ have weak  dependence on $U$, and increasing $U$ improves $d_{11}$ (absolute value from 2.85 pm/V to 7.12 pm/V) due to enhanced $e_{11}$ (absolute value). It is found that the $d_{31}$ is very small (-0.125 pm/V to -0.087 pm/V with increasing $U$).
At a typical $U$=2.0 eV, the calculated $d_{11}$ is -5.61 pm/V, which  is higher than or  compared with ones of  other 2D known materials\cite{f2,f12}.
So, YBrI monolayer is a PFM.
\begin{figure*}
   \includegraphics[width=15cm]{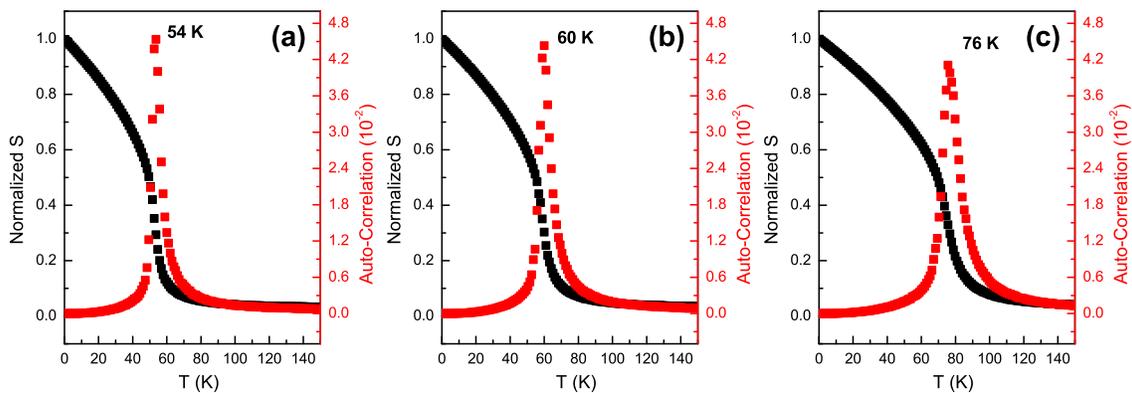}
  \caption{(Color online)For YBrI monolayer, the normalized magnetic moment (S) and auto-correlation  as a function of temperature with $U$$=$0.0 eV (a), 2.0 eV (b) and $a/a_0$=0.94 (c) at $U$$=$2.0 eV.}\label{tc}
\end{figure*}

\section{Strain  effects}
The electronic, piezoelectric, valley and topological properties of 2D materials can be easily tuned by strain engineering\cite{m1,m2,m3,m4}.
We use $a/a_0$ (0.94-1.06) to simulate biaxial strain, in which  $a$/$a_0$  denotes   strained/unstrained lattice constant.
  The  $a/a_0$$<$1  indicates compressive  strain, while  the $a/a_0$$>$1  means tensile strain.
Taking a typical $U$=2.0 eV,  the strain effects on physical properties of YBrI are investigated.

\autoref{s-ae} shows  the energy difference between  AFM and FM ordering as a function of $a/a_0$, and all strained YBrI are  FM ground state in considered strain range. It is found that compressive strain can enhance FM interaction,  which is in favour of high Curie temperature.
Within SOC,  the strained  energy band structures with both out-of-plane and in-plane magnetic anisotropy are plotted in FIG.4 of ESI, and the corresponding gaps and valley splitting are shown in \autoref{s-gap}. With $a/a_0$ from 0.94 to 1.06,  YBrI is always a FM semiconductor for both out-of-plane and in-plane magnetic anisotropy, and the gap firstly increases, and then decreases.
For out-of-plane case, the tensile strain can enhance valley splitting.  According to \autoref{s-ae}, the inherent easy magnetization axis of YBrI is always  in-plane in considered strain range, which means that strained YBrI is a common FM semiconductor, not a FV material.
So, magnetic and electronic properties of YBrI monolayer are robust against strain.

Next,  the strain influences on  piezoelectric properties of YBrI are investigated. The elastic constants  ($C_{11}$, $C_{12}$,  $C_{11}$-$C_{12}$ and $C_{11}$+$C_{12}$) and  piezoelectric  stress  coefficients  ($e_{11}$ and $e_{31}$) along  the ionic  and electronic contributions  vs $a/a_0$ are plotted  in FIG.5 and FIG.6 of ESI. The $d_{11}$/$d_{31}$ as a function of $a/a_0$ are plotted in \autoref{s-d11}.
It is found that increasing tensile strain can improve $d_{11}$  (absolute value) due to decreased $C_{11}$-$C_{12}$  and enhanced $e_{11}$ based on \autoref{pe2}. At 1.06 strain, the $d_{11}$ (-11.14 pm/V) is about two times  as large as  unstrained one (-5.61 pm/V).
 However,  the  strain has weak  influence on $d_{31}$.  The strained YBrI monolayer is mechanically stable, because calculated elastic constants (FIG.5 of ESI)  satisfy  the mechanical stability criteria. So, strain is an effective way to tune piezoelectric properties of YBrI.

\section{Curie temperature}
 Increasing electronic correlation  and compressive strain can enhance  the strength of  FM interaction, which can improve Curie temperature $T_C$ of monolayer YBrI. To confirm this, we estimate  $T_C$  at  $U$$=$0.0 eV and 2.0 eV,  and at $a/a_0$$=$0.94 with $U$$=$2.0 eV by MC  simulation.
The Heisenberg spin Hamiltonian can be written as:
  \begin{equation}\label{pe0-1-1}
H=-J\sum_{i,j}S_i\cdot S_j-A\sum_i(S_i^z)^2
 \end{equation}
in which   $S_i$ and $S_j$    are   the
the spin
operators on sites $i$ and $j$, and  $S_i^z$ represents the spin orientation along the $z$
direction, and $J$ and  $A$ are  the nearest neighbor exchange parameter and   MAE.

The $J$ with normalized spin vector ($|S|$=1) is determined from the energy
difference between  AFM  ($E_{AFM}$) and FM ($E_{FM}$)  with rectangle supercell.
 Based on the FM
and AFM configurations, $J$  can be
obtained by equations:
\begin{equation}\label{pe0-1-2}
E_{FM}=E_0-6J-2A
 \end{equation}
  \begin{equation}\label{pe0-1-3}
E_{AFM}=E_0+2J-2A
 \end{equation}
where $E_0$ is the total energy of systems without magnetic coupling.
The  corresponding $J$ can be attained:
  \begin{equation}\label{pe0-1-3}
J=\frac{E_{AFM}-E_{FM}}{8}
 \end{equation}
The calculated normalized $J$  are  4.13 meV and 4.70 meV for $U$$=$0.0 eV and 2.0 eV, and is 6.10 meV  at $a/a_0$$=$0.94 with $U$$=$2.0 eV.
 The  normalized magnetic moment and auto-correlation  as a function of  temperature at  representative $U$ (0.0 eV and 2.0 eV)  and $a/a_0$ (0.94 with $U$$=$2.0 eV) values are plotted in \autoref{tc}, and the predicted $T_C$ is about 54 K, 60 K and 76 K, respectively. So, increasing electronic correlation  and compressive strain indeed can improve $T_C$ of YBrI. The predicted $T_C$ (60 K) of YBrI is very lower than those of  $\mathrm{YBr_2}$ (220 K) and  $\mathrm{YI_2}$ (230 K)\cite{f9}. The possible reason of  huge difference is that the Ising model is adopted in Ref.\cite{f9}. The predicted $T_C$ of YBrI is higher than those of  experimental discovered 2D ferromagnetic
materials $\mathrm{CrI_3}$ (45 K) and $\mathrm{Cr_2Ge_2Te_6}$ (30 K)\cite{f3-2,f3-3}.

\section{Discussion and Conclusion}
Monolayer  $\mathrm{YM_2}$ (M=I, Br and Cl) have been predicted to be  dynamically
and thermally stable 2D FM semiconductors\cite{f9}. Based on these monolayers, Janus YClBr and YClI monolayers can also be constructed, which should be FM and  piezoelectric. Except for piezoelectric properties, the main analysis and results about YBrI  can  be readily extended to monolayer YMN (M/N=Cl, Br and I).  Monolayer $\mathrm{YM_2}$ (M=I, Br and Cl) possess a reflection symmetry with respect to the central Y
atomic plane, which  requires that $e_{31}$/$d_{31}$=0. Their piezoelectric polarizations are confined along the in-plane armchair direction.
However, for monolayer YMN (M/N=Cl, Br and I; M$\neq$N), both $e_{11}$/$d_{11}$ and $e_{31}$/$d_{31}$ are nonzero, which means that both in-plane (along the armchair direction) and
vertical piezoelectric polarizations are allowed. For monolayer FeMN (M/N=F, Cl, Br and I)\cite{f11,f11-1},
electronic correlation  can induce many novel electronic states and topological phase transition. However, qualitative results should be invariable with varied $U$ for monolayer YMN (M/N=Cl, Br and I).

In summary, based on reliable DFT calculations, we systemically investigate  the stability,  and the electronic, magnetic and piezoelectric
properties of Janus YBrI monolayer. For out-of-plane magnetic anisotropy,  the spontaneous valley polarization can be observed with valley splitting being larger than 82 meV in considered $U$ range. However, for in-plane  magnetic anisotropy, YBrI is a common FM semiconductor without spontaneous valley polarization.  The MAE calculations show that the intrinsic easy axis of YBrI is always in-plane with $U$ from 0.0 eV to 3.0 eV.
Both  in-plane and  out-of-plane piezoelectric polarizations can be observed due to broken  inversion and mirror symmetry,
and the  predicted in-plane $d_{11}$   is  higher than/compared with ones of many familiar 2D materials.  The predicted $T_C$ is larger than the reported values for 2D  $\mathrm{CrI_3}$  and $\mathrm{Cr_2Ge_2Te_6}$\cite{f3-2,f3-3}.
It is found that  strain engineering is very effective way to tune physical properties of YBrI.
Our findings suggest that YBrI can have
promising applications in  multipurpose spintronic devices.

~~~~\\
~~~~\\
\textbf{Conflicts of interest}
\\
There are no conflicts to declare.

\begin{acknowledgments}
This work is supported by Natural Science Basis Research Plan in Shaanxi Province of China  (2021JM-456),  the Nature Science Foundation of China (Grant No.11974393) and the Strategic Priority Research Program of the Chinese Academy of Sciences (Grant No. XDB33020100). We are grateful to the Advanced Analysis and Computation Center of China University of Mining and Technology (CUMT) for the award of CPU hours and WIEN2k/VASP software to accomplish this work.
\end{acknowledgments}

\end{document}